\documentclass[amsmath,superscriptaddress,prb,twocolumn] {revtex4}
\usepackage{bm}
\usepackage{enumerate}
\usepackage{graphicx}
\usepackage[dvips]{epsfig}
\usepackage{epsf}
\usepackage{xcolor}

\makeatletter
\newcommand{\Rmnum}[1]{\expandafter\@slowromancap\romannumeral #1@}
\makeatletter

\begin{document}
\title{Ferromagnet with a noncollinear antiferromagnetic order and anomalous Hall effect}
\author{Vladimir A. Zyuzin}
\affiliation{L.D. Landau Institute for Theoretical Physics, 142432, Chernogolovka, Russia}
\begin{abstract}
In this paper we introduce a theoretical model of a metallic magnetic system with noncollinear antiferromagnetic order. We introduce a mechanism of indirect interaction of conducting fermions with localized spins based on the tunneling processes of conducting fermions through the localized spins. We demonstrate that interaction of conducting fermions with noncollinear antiferromagnetic order results in odd in momentum spin-momentum locking. The interaction resembles Rashba spin-orbit coupling but breaks the time-reversal symmetry. As a result, we show that a ferromagnet with the noncollinear antiferromagnetic order is an insulator with anomalous Hall effect occurring without any spin-orbit coupling.
\end{abstract}
\maketitle

\section{Introduction}

Anomalous Hall effect is a unique property of metallic ferromagnets \cite{KarplusLuttinger} in which Hall voltage appears without the external magnetic field. The mechanism is based on an interplay of the ferromagnetic exchange interaction generated spin-splitting of the conducting fermions and finite spin-orbit coupling \cite{AHE_RMP}. These two interactions result in an anomalous velocity \cite{BerryReview} of conducting fermions which can be understood to be due to the effective magnetic field acting in ${\bf k}-$space, the Berry curvature \cite{Berry1984}.

Metallic collinear antiferromagnets can show anomalous Hall effect despite of the fully compensated magnetic N\'{e}el order. 
 Antiferromagnets were classified for all symmetry classes in \cite{Turov1965}. The classification is based on constructing what can be called as the Dzyaloshinskii's invariant \cite{Dzyaloshinskii1958,Turov1965,GolubinskiiZyuzin} for the existence of finite magnetic moment in collinear N\'{e}el order antiferromagnets \cite{Dzyaloshinskii1958}. To realize the Dzyaloshinskii's invariant, certain crystal symmetries must be broken in the system.
In \cite{TurovShavrov,Turov1990,AHE_AFM,GolubinskiiZyuzin} it has been shown that antiferromagnets with certain symmetries between the magnetic sublattices may show the anomalous Hall effect. These are called as the weak ferromagnets \cite{Dzyaloshinskii1958,Turov1965,GolubinskiiZyuzin}. Antiferromagnets with compensated N\'{e}el order but in which there is no symmetry between the magnetic sublattices also show the anomalous Hall effect \cite{Naka2020,Zyuzin2024,Zyuzin2025a}. Such antiferromagnets are called as ferrimagnets \cite{Turov1965}. The mechanism of the anomalous Hall effect in both cases is the interplay of momentum-dependent exchange interaction of conducting fermions due to the interaction of fermions with the N\'{e}el order \cite{Noda2016,Okugawa2018,HayamiYanagiKusunose2019,AHE_AFM,Ahn2019,Rashba2020,Naka2020,HayamiYanagiKusunose2020,EgorovEvarestov2021,Brekke2023,AgterbergPRB2024,Zyuzin2024,Zyuzin2025a,Antonenko2025,Mineev2025a,Mineev2025b,GolubinskiiZyuzin} and symmetry allowed intrinsic spin-orbit coupling\cite{GolubinskiiZyuzin}. In \cite{GolubinskiiZyuzin} it has been shown that both necessary ingredients for the non-zero anomalous Hall effect in antiferromagnets originate from the broken symmetries that allowed for the Dzyaloshinskii's invariant\cite{Turov1965} in the system.

\begin{figure}[b] 
\begin{tabular}{cc}
\includegraphics[width=0.9 \columnwidth ]{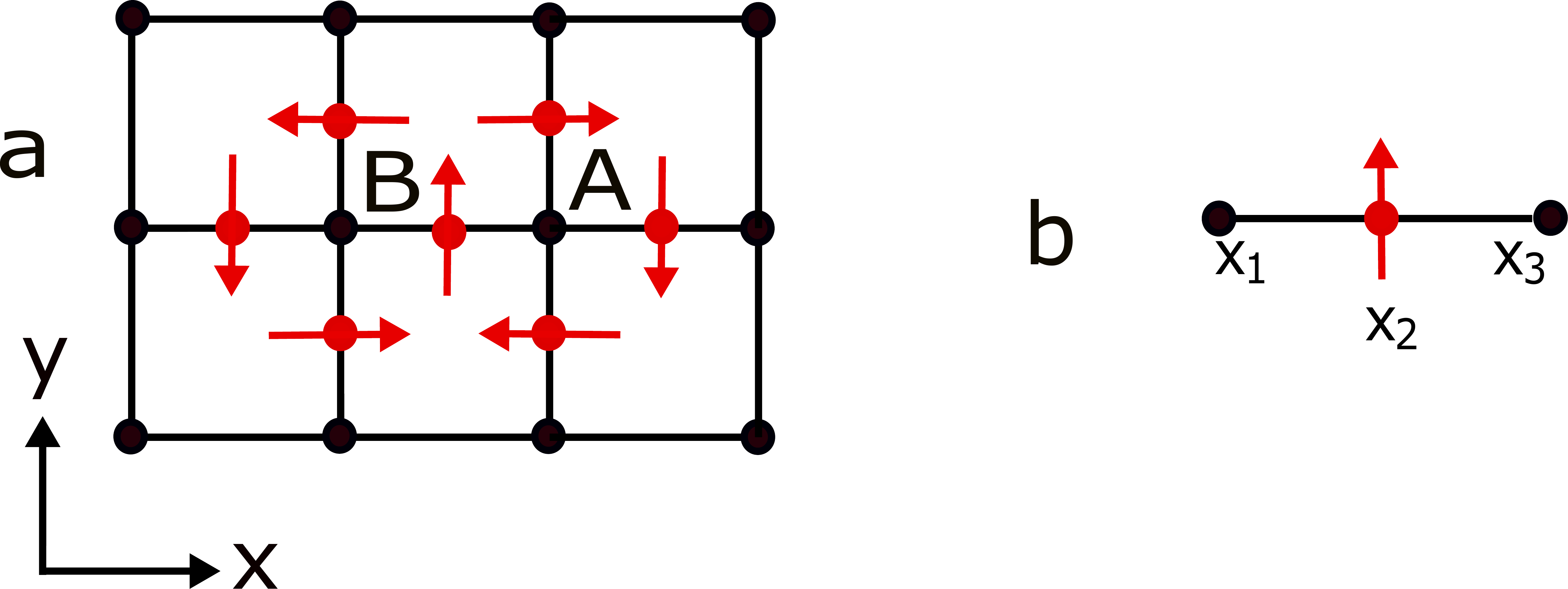} ~~~~ 
\end{tabular}

\protect\caption{(a): a model of a noncollinear antiferromagnetic order. Black circles are the sites conducting fermions reside on. Red circles are sites with a localized spin. Direction of the spin on the red site is set by the red arrow. There is a tunneling of conducting fermions between the black sites.  Fermion energy at the red site is higher than the one on the black. Thus, there is a tunneling of conducting fermions through the red sites which results in the noncollinear antiferromagnetic order interaction given in Eq. {\ref{antitoroidal}}. (b): a model of three sites on a line used as an example where fermion's momentum-dependent exchange interaction with localized spin appears.}
\label{fig:fig1}  
\end{figure}

Here we suggest a scenario of appearence of the anomalous Hall effect in magnetic systems with noncollinear antiferromagnetic order without any spin-orbit coupling. 
In Refs. \onlinecite{Noda2016,Okugawa2018,HayamiYanagiKusunose2019,AHE_AFM,Ahn2019,Rashba2020,Naka2020,
HayamiYanagiKusunose2020,EgorovEvarestov2021,Brekke2023,AgterbergPRB2024,Zyuzin2024,Zyuzin2025a,
Antonenko2025,Mineev2025a,Mineev2025b,GolubinskiiZyuzin}
 momentum-dependent exchange interaction of conducting fermions occurring in metallic collinear antiferromagnets was shown to be due to anisotropic fermion tunneling within the magnetic sublattices. This was achieved by different local environment of the magnetic sublattices. Here we introduce noncollinear antiferromagnetic order as an example of another type of momentum-dependent exchange interaction of conducting fermions with localized spin. In particular, this is the situation when conducting fermions and localized spins reside on different sites and the interaction occurs during the tunneling processes of the former through the sites with localized spin. We consider noncollinear antiferromagnetic order on the square lattice shown in Fig. (\ref{fig:fig1}a) as an example where such momentum-dependent interaction occurs.
We show that interaction of the conducting fermions with the noncollinear antiferromagnetic order results in effective Rashba spin-orbit coupling \cite{Vas'ko,BychkovRashba,Dyakonov} which breaks the time-reversal symmetry. 
We show that making the system ferromagnetic will allow for the anomalous Hall effect without any spin-orbit coupling.
noncollinear antiferromagnetic orders have been extensively studied in the past \cite{Turov1965,BorovikRomanov1988,Mineev1996,ChenNiuMacDonaldPRL2014,Hellens2023,Sukhachov2024, Agterberg2025,Liu2022,Fukami2025,Parkin2025}. 
The obtained in this paper time-reversal symmetry breaking effective Rashba spin-orbit coupling is closely related to recently proposed magnets with \textit{p-}wave spin-splitting and other unusual metals with spin ordering \cite{ChenNiuMacDonaldPRL2014,Liu2022,Fukami2025,Parkin2025,Hellens2023,Sukhachov2024,Agterberg2025,Mineev2025a,Mineev2025b}, but to the best of our knowledge this exact form and its consequences on physical properties haven't been discussed anywhere before.

\section{A model of a noncollinear antiferromagnetic order}

We will be studying a system shown in
Fig. (\ref{fig:fig1}a). There the black sites corrspond to the sites conducting fermions reside on, while red sites are where localized spins are located. 
We assume that the localized spins form noncollinear antiferromagnetic order as shown by red arrows in Fig. (\ref{fig:fig1}a). The purpose of this Section is to derive an effective exchange interaction of conducting fermions with localized spins. With this we omit the Heisenberg exchange Hamiltonian between localized spins and assume that the noncollinear antiferromagnetic order shown in Fig. (\ref{fig:fig1}a) has been stabilized. The question of stabilization of such noncollinear orders has been extensively studied, for example in Refs. (\onlinecite{Turov1965,Mineev1996,Agterberg2025}) and many more. We also assume that the order is fixed and isn't fluctuating.
As shown in Fig. (\ref{fig:fig1}a) the vorticity of localized spins on one plaquette is opposite to the vorticity on the neighboring.
Because of that there are two sites, $\mathrm{A}$ and $\mathrm{B}$, in the unit cell of conducting fermions. 
The noncollinear antiferromagnetic order is similar to the conventional collinear antiferromagnetic systems in a sense that in both the total spin is zero but the orders break the time-reversal symmetry. 
We note that in the studied model of Ref. \onlinecite{Hellens2023} there is no antiferromagnetic symmetry in the system.
Since the conducting fermions and the localized spins are located at different sites, then
there is no direct exchange interaction of conducting fermions with the localized spins in the form of a on-site Zeeman-field. 
However, there might be an interaction during the tunneling processes of the conducting fermions through the sites with localized spins. In Ref. \onlinecite{Sukhachov2024} similar noncollinear antiferromagnetic order was studied but in their case conducting fermions and localized spins were on the same site.  
In the basis of sublattices and spins $\hat{\Psi} = \left( \Psi_{\mathrm{A};\uparrow;{\bf k}}, \Psi_{\mathrm{A};\downarrow;{\bf k}}, \Psi_{\mathrm{B};\uparrow;{\bf k}}, \Psi_{\mathrm{B};\downarrow;{\bf k}}\right) $ a
Hamiltonian describing an interaction of conducting fermions with the noncollinear antiferromagnetic order shown in Fig. (\ref{fig:fig1}a) is
\begin{align}\label{antitoroidal}
\hat{H}_{\mathrm{nc}} = 
\left[\begin{array}{cc} 0 &   i\gamma(\sigma_{x}s_{y} - \sigma_{y}s_{x})  \\
- i\gamma(\sigma_{x}s_{y} - \sigma_{y}s_{x})  & 0 \end{array}\right]
\end{align}
where $s_{x/y}=\sin(k_{x/y})$ is introduced for brevity, $\gamma$ is a constant defining interaction of conducting fermions with the noncollinear antiferromagnetic order, and $\sigma$ are Pauli matrices acting in the spin space. Although this term appears almost like the Rashba spin-orbit coupling \cite{Vas'ko,BychkovRashba,Dyakonov}, it nevertheless breaks the time-reversal symmetry. The difference of the noncollinear antiferromagnetic order with the Rashba spin-orbit coupling is in the extra factor of $i$ in the former. Because of that it can only appear in systems with at least two sublattices. For example, Rashba spin-orbit coupling would appear in the system if a reflection in the $x-y$ plane is broken by external electric field applied in the perpendicular direction to the plane of the system.
In this case 
\begin{align}\label{Rashba}
\hat{H}_{\mathrm{Rashba}} = \lambda \left[\sigma_{x}\sin(k_{y}) - \sigma_{y}\sin(k_{x}) \right] \tau_{1} ,
\end{align}
where $\lambda$ is a constant and $\tau_{1}$ is the Pauli matrix acting in the sublattice space, will appear in the Hamiltonian. It can be seen this term doesn't break the time-reversal symmetry. We will not consider regular Rashba spin-orbit coupling in what follows.

Let us derive interaction Eq. (\ref{antitoroidal}) with the noncollinear antiferromagnetic order from the scratch. 
Consider three sites on a line shown in Fig. (\ref{fig:fig1}b). 
Fermions reside on sites at $x=x_{1/3}$, and these sites are with fermion energy $E_{0}$. 
The site at $x=x_{2}$ has fermion energy larger than that of the first two, $E_{1}>E_{0}$. 
In addition, $x=x_{2}$ site has a localized spin pointing in, for example, $y$ direction. 
There is a tunneling $t$ of fermions between sites $x_{1/3}$ and $x_{2}$.
Hamiltonian of the system can be written in the form
\begin{align}\label{3}
H_{3} &= \sum_{i=1,3;\alpha} E_{0}\psi^{\dag}_{\alpha}(x_{i})\psi_{\alpha}(x_{i})
\\
&
+
\sum_{\alpha\beta}\left[E_{1}\delta_{\alpha\beta}+m_{y} (\sigma_{y})_{\alpha\beta}\right] c^{\dag}_{\alpha}(x_{2})c_{\beta}(x_{2})
\nonumber
\\
&
+t\sum_{i=1,3;\alpha}\psi^{\dag}_{\alpha}(x_{i})c_{\alpha}(x_{2})+t\sum_{i=1,3;\alpha}c_{\alpha}^{\dag}(x_{2})\psi_{\alpha}(x_{i}),
\nonumber
\end{align}
where $\psi^{\dag}$, $\psi$, $c^{\dag}$ and $c$ are fermion operators, $\alpha$ and $\beta$ denote spins, and $m_{y}$ is the exchange interaction on site $x=x_{2}$ The names are chosen to highlight different fermion species, namely, operators $\psi^{\dag}$ and $\psi$ act on fermions on the $x=x_{1/3}$ sites, while $c^{\dag}$ and $c$ on those on the $x=x_{2}$ site.
We integrate the fermion state at $x=x_{2}$ site out and obtain tunneling of fermions between $x_{1}$ and $x_{3}$ sites to the lowest order in $t$,
\begin{align}\label{perturbation}
H_{1\rightarrow3} = t^2\sum_{\alpha\beta} \psi^{\dag}_{\alpha}(x_{3})\psi_{\beta}(x_{1})\frac{[(E_{0}-E_{1})\delta_{\alpha\beta}-m_{y} (\sigma_{y})_{\alpha\beta}]}{(E_{0} - E_{1})^2 - m_{y}^2},
\end{align}
in which a spin structure $\propto (\sigma_{y})_{\alpha\beta}$ appears. 
We conclude that the exchange interaction of fermions of the first type located on sites $x=x_{1/3}$ with a localized spin located on the site $x=x_{2}$ occurs during the tunneling processes.

Now let us apply this knowledge to the lattice with noncollinear antiferromagnetic order shown in Fig. (\ref{fig:fig1}a). There the conducting fermions reside on black sites and localized spins that form the noncollinear antiferromagnetic order are on the red sites.
Then, summing up the two-step tunneling processes between the $\mathrm{B}$ and $\mathrm{A}$ sublattices through the sites with the localized spins, we get 
\begin{align}
\hat{H}^{\mathrm{B} \rightarrow \mathrm{A}}_{\mathrm{nc}} & =\frac{\gamma}{2}\int_{\bf k} \psi^{\dag}_{\mathrm{A};{\bf k};\alpha}\psi_{\mathrm{B};{\bf k};\alpha} \left( -e^{ik_{x}}\sigma_{y} + e^{-ik_{x}}\sigma_{y}   \right)
\nonumber
\\
&
+\frac{\gamma}{2}\int_{\bf k} \psi^{\dag}_{\mathrm{A};{\bf k};\alpha}\psi_{\mathrm{B};{\bf k};\alpha} 
\left( e^{ik_{y}}\sigma_{x} - e^{-ik_{y}}\sigma_{x}   \right)
\nonumber
\\
&
=i\gamma \int_{\bf k} \psi^{\dag}_{\mathrm{A};{\bf k};\alpha}\psi_{\mathrm{B};{\bf k};\alpha} \left[ \sin(k_{x})\sigma_{y}- \sin(k_{y})\sigma_{x}\right].\label{tight13}
\end{align} 
In this way the Hamiltonian Eq. (\ref{antitoroidal}) is justified.
There is no on-site spin-dependent term, i.e. of the $\psi^{\dag}_{\alpha}(x_{1/3})\psi_{\beta}(x_{1/3}) (\sigma_{y})_{\alpha\beta}$ structure, appearing in any order of the perturbation theory because of the opposite sign of $\sigma_{y}$ term on the neighboring links (shown in Fig. (\ref{fig:fig1}a)). Such terms sum up to zero.

\begin{figure}[h] 
\begin{tabular}{cc}
\includegraphics[width=0.9 \columnwidth ]{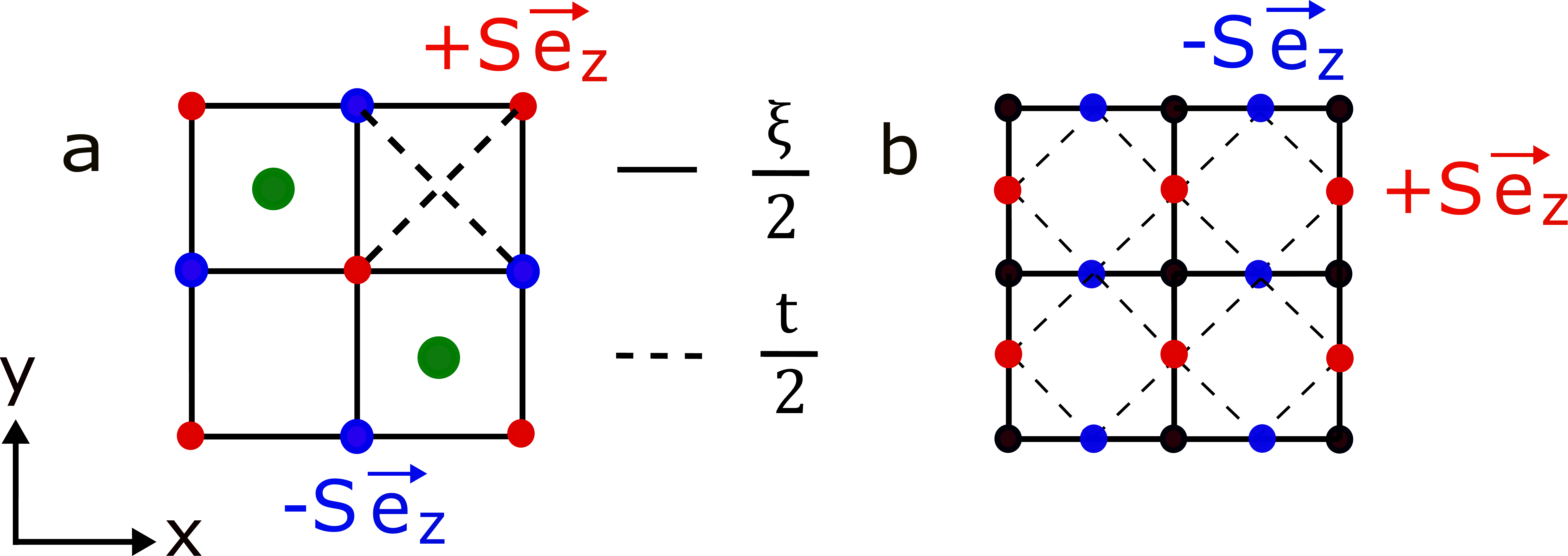} ~~~~ 
\end{tabular}

\protect\caption{Two different models of genuine N\'{e}el ordered antiferromagnets on a checkerboard lattice.
Red and blue sites correspond to localized fermions which form the N\'{e}el order (in $z-$direction, for example), in which spins on the red/blue sites point in up/down in $z-$direction. Conducting fermions in the model (a) reside on the colored sites, while in the model (b) they reside on black sites. The green atom shown in (a) is non-magnetic and it is assumed it blocks fermion tunneling through it. Hence, there is an anisotropic tunneling processes of conducting fermions described by the dashed lines in the model (a) which result in the $d-$wave spin-splitting described in Eq. (\ref{checker2}). In the model (b) the dashed lines describe Heisenberg exchange interaction between localized spins. Direct conducting fermion tunneling in the model (b) is between black sites, while there are also two-step tunneling processes through the colored sites in correspondence with Fig. (\ref{fig:fig1}b). The latter processes result in the $d-$wave spin-splitting of conducting fermions described in Eq. (\ref{dwave}).}
\label{fig:fig11}  
\end{figure}

It is instructive to outline the mechanism of another type of the momentum-dependent exchange interaction occurring in antiferromagnets. Let us consider a model of genuine (with zero-net magnetization) antiferromagnet on the checkerboard lattice shown in Fig. (\ref{fig:fig11}a). There conducting fermions and localized spins which form the N\'{e}el order are on the same site. Therefore, there is a direct on-site exchange interaction of the conducting fermions with the localized spins written as $\pm {\bf m}\cdot{\bm \sigma}$. We then write a Hamiltonian of conducting fermions
\begin{align}\label{checker}
\hat{H}_{\mathrm{checker}} = 
\left[\begin{array}{cc} {\bf m}\cdot{\bm \sigma} - t_{\bf k} & \xi_{\bf k} \\ 
\xi_{\bf k} & -{\bf m}\cdot{\bm \sigma} + t_{\bf k} \end{array} \right],
\end{align}
where in addition, there is a regular tunneling between sites given by $\xi_{\bf k} = \xi \left[\cos(k_{x})+\cos(k_{y})\right]$, and anisotropic second-nearest neighbor tunneling $t_{\bf k}$. The green atom placed in the center of the square plaquette as shown in Fig. (\ref{fig:fig11}) is assumed to block the fermion tunneling through it. Therefore, for the fermion tunneling from red to red we may write $t^{\mathrm{R}}_{\bf k} = t\cos(k_{x}+k_{y})$ and from blue to blue it is $t^{\mathrm{R}}_{\bf k} = t\cos(k_{x}-k_{y}) $. It is the anisotropic part of these two processes defines $t_{\bf k}$ in Eq. (\ref{checker}), namely $t_{\bf k} = t\sin(k_{x})\sin(k_{y})$. There are many ways to understand the momentum-dependent spin-spitting of conducting fermions in the system. One may square the Hamiltonian to obtain 
\begin{align}\label{checker2}
\hat{H}_{\mathrm{checker}}^2 = \xi_{\bf k}^2 + m_{z}^2 + t_{\bf k}^2 
-2m_{z} \sigma_{z} t_{\bf k},
\end{align}
and hence the momentum-dependent spin-splitting $\propto -2 m_{z} \sigma_{z} t \sin_{k_{x}}\sin(k_{y})$. 
Such a mechanism has received much of attention recently. For example see Refs. (\onlinecite{Noda2016,Okugawa2018,HayamiYanagiKusunose2019,AHE_AFM, Ahn2019,Rashba2020,Naka2020,HayamiYanagiKusunose2020,EgorovEvarestov2021,Brekke2023, AgterbergPRB2024,Zyuzin2024,Zyuzin2025a,Antonenko2025,Mineev2025a,Mineev2025b,GolubinskiiZyuzin}) and many more.
Let us now consider a model of genuine N\'{e}el ordered antiferromagnet on another checkerboard lattice shown in Fig. (\ref{fig:fig11}b).
Now, the conducting fermions reside on black sites, while the N\'{e}el is on the colored sites. There is no direct on-site exchange interaction between the conducting fermions and the N\'{e}el order. However, there may be two-step fermion tunneling between the black sites through the colored ones. 
Spin-dependent part of such tunneling processes is derived to be
\begin{align}\label{dwave}
\hat{H}_{\mathrm{d-wave}} = \gamma_{\mathrm{dw}}[\cos(k_{x})-\cos(k_{y})]\hat{\sigma}_{z},
\end{align}
where $\gamma_{\mathrm{dw}}$ is a constant. We note that in contrast with the first example of the N\'{e}el order on the checkerboard lattice Eq. (\ref{checker}), the unit cell in the system shown in Fig. (\ref{fig:fig11}) contains only one black site despite of the two sublattices of the N\'{e}el order.
All in all, we have demonstrated how the $d-$wave spin-splitting of conducting fermions in case of absent direct exchange interaction with the N\'{e}el order appears in genuine antiferromagnets. Antiferromagnets with such spin-splitting are known to show inetersting, typically anisotropic and N\'{e}el vector dependent, effects reviewed in Ref. \onlinecite{Turov1990}.
Among them we would like to mention quadratic in magnetic field Faraday effect \cite{KharchenkoBibikEremenko1985,EremenkoKharchenko1987} experimentally observed in a weak ferromagnet CoF$_2$ \cite{KharchenkoBibikEremenko1985,EremenkoKharchenko1987}, d-wave Hall effect (which is related to the mentioned Faraday effect) and linear magnetoconductivity both predicted in \cite{VorobevZyuzin2024}, spin-splitter effect \cite{Naka2019}, and others. It should be noted that the weak ferromagnet CoF$_2$ has a rutile type crystal structure, and, thus, has finite magnetization for certain directions of the N\'{e}el vector in accord with table 3 in Ref. (\onlinecite{Turov1965}). The experiment\cite{KharchenkoBibikEremenko1985} was conducted for the directions of the N\'{e}el vector for which the magnetization is zero. 

All in all, we may claim that our derivations Eqs. (\ref{antitoroidal}), (\ref{tight13}), (\ref{dwave}) are consistent with known results, and we proceed with studying the anomalous Hall effect in the system shown in Fig. (\ref{fig:fig1}).

\begin{figure}[h] 
\includegraphics[width=0.45 \columnwidth ]{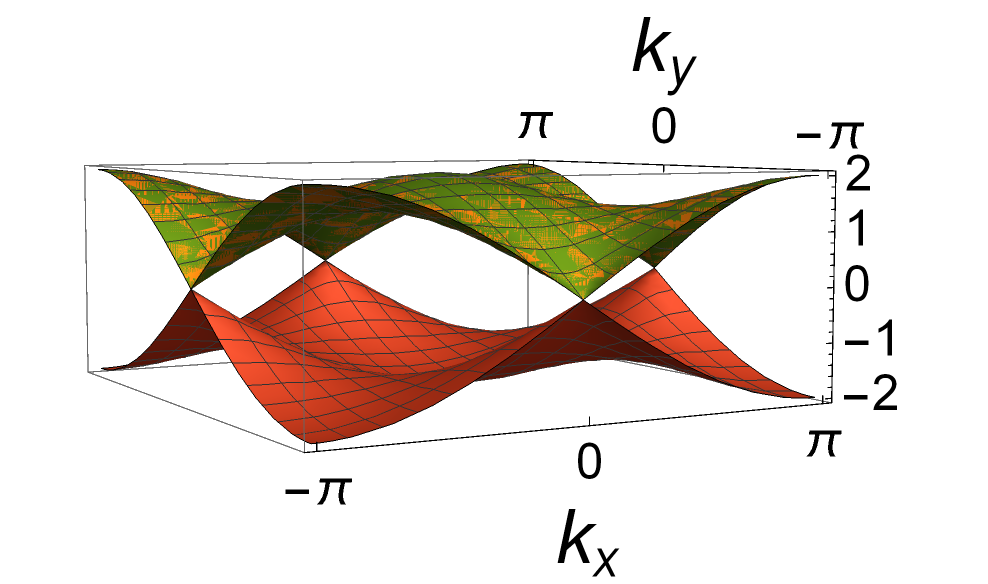} 
\includegraphics[width=0.45 \columnwidth ]{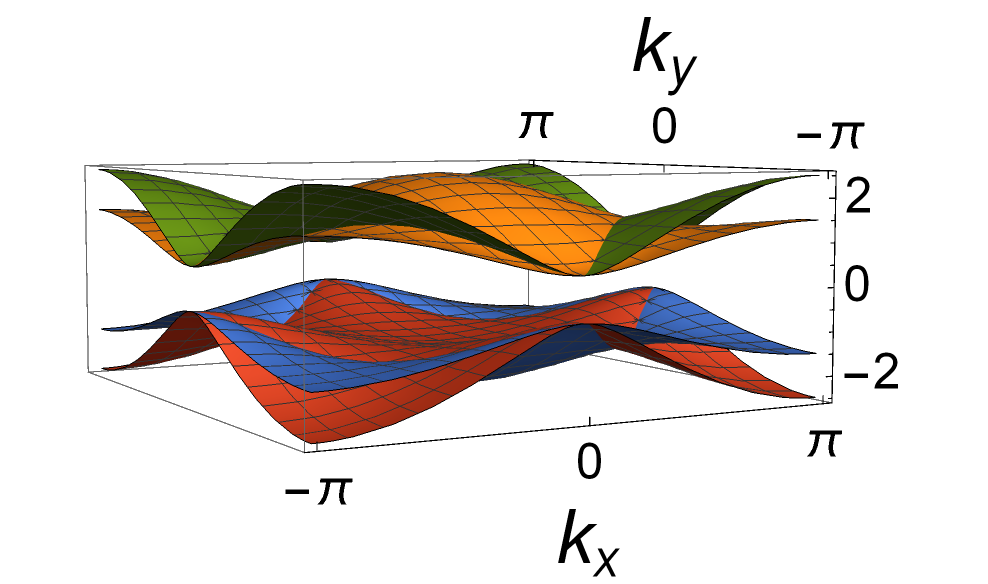} 

\protect\caption{Spectrum of the system with noncollinear antiferromagnetic order described by Hamiltonian $\hat{H} = \hat{H}_{\mathrm{nc}}+\hat{H}_{\mathrm{fm}}$ defined in Eq. (\ref{antitoroidal}) and Eq. (\ref{fm}). Left plot is for $h_{z} = 0$ in which the conduction and valence bands touch at four points. Right: finite magnetic field, $h_{z} = 0.5\xi$, opens up a gap at these points. Both plots are for $\gamma =\xi$ in order to highlight gap opening. Physical prameters $h_{z}$ and $\gamma$ are in units of $\xi$, such that $\xi=1$.}
\label{fig:fig2}  
\end{figure}

\section{A model of a ferromagnet with the noncollinear antiferromagnetic order}
Having understood the nature of momentum-dependent exchange interaction of conducting fermions with the noncollinear antiferromagnetic order shown in Fig. (\ref{fig:fig1}), we now wish to understand its consequences on the physical properties.
A combination of time-reversal and translation opeartions is the symmetry of the system. In addition, a combination of a mirror reflection in the $x-y$ plane and time reversal is also a symmetry of the system. Both symmetries forbid existence of the magnetic moment in the system. Hence, not much we can anticipate in terms of the anomalous transport properties of conducting fermions.
In addition to the interaction of conducting fermions with the noncollinear antiferromagnetic order, we assume that there is on-site ferromagnetism pointing in $z-$direction. Now, both symmetries are explicitly broken, and we expect anomalous Hall effect to occur in the system.
The tight-binding Hamiltonian of conducting fermions is now
\begin{align}\label{fm}
\hat{H} = 
\left[\begin{array}{cc} h_{z}\sigma_{z} - \mu & \xi_{{\bf k}} + i\gamma(\sigma_{x}s_{y} - \sigma_{y}s_{x}) \\
\xi_{{\bf k}} - i\gamma(\sigma_{x}s_{y} - \sigma_{y}s_{x}) & h_{z}\sigma_{z} - \mu \end{array}\right],
\end{align}
where $\xi_{\bf k} = \xi [\cos(k_{x})+\cos(k_{y})]$ is the nearest-neighbor tunneling of conducting fermions, $\mu$ is the Fermi energy, and $h_{z}$ is the ferromagnetic exchange interaction due to the ferromagnetic order in $z-$direction. 

No other momentum and spin dependent interaction will appear upon adding the on-site $ \psi^{\dag}_{\alpha}(x_{i})\psi_{\beta}(x_{i})h_{z}(\sigma_{z})_{\alpha\beta}$ to the Eq. (\ref{3}) and performing the perturbation theory scheme discussed in Eq. (\ref{perturbation}). One may think that genuine Rashba-like spin-orbit coupling of the form given in Eq. (\ref{Rashba}) with the Rashba constant proportional to $\lambda \propto t^2h_{z}m_{x/y}$ can be generated in the perturbation theory. Indeed, this combination of physical parameters respects the time-reversal symmetry operation. However, it will vanish because a combination of reflection in the $x-y$ plane and translation is the symmetry of the lattice. Recall that all spins transform as pseudovectors. Therefore, in-plane spins flip their direction upon the reflection, and subsequent translation restores the system. Normal to the plane ferromagnetic moment doesn't change under the operation of the $x-y$ plane mirror reflection. This is inuitively correct because ferromagnetism can't produce electric field in the settings and assumptions of the studied system. In other words, by the construction of the system there are no dipole moments ferromagnetism can result in. 

We wish to find eigenvalues and eigenvectors of the $\hat{H}\Psi_{\bf k} = \epsilon_{\bf k} \Psi_{\bf k}$ equation. The first part is straightforward, $\epsilon_{{\bf k};1;\pm} + \mu = \pm  \sqrt{(h_{z} + \xi_{\bf k})^2 + \gamma^2 s_{\bf k}^2}$ and $\epsilon_{{\bf k};2;\pm} + \mu = \pm  \sqrt{(h_{z} - \xi_{\bf k})^2 + \gamma^2 s_{\bf k}^2}$ are the eigenvalues, where $s_{\bf k}^2 = s_{x}^2+s_{y}^2$ was introduced. The spectrum for $h_{z} = 0$ is shown in the left of Fig. (\ref{fig:fig2}).
Conduction and valence bands touch at four points in the Brillouin zone (BZ). 
Indeed, by setting, for example $k_{x} = 0$ and $k_{y} = \pi$ we get both $s_{\bf k} = 0$ and $\xi_{\bf k} = 0$.
In addition, the spectrum at $h_{z} = 0$ is degenerate which corresponds to the symmetry between the sublattices which is a combination of time-reversal and translation. In terms of the Hamiltonian, the symmetry between the sublattices translates in to a $\hat{H} = \tau_{1}\hat{\Theta}\hat{H}\hat{\Theta}\tau_{1}$ symmetry relation, where $\tau_{1}$ is the Pauli matrix acting in the sublattice space, and $\hat{\Theta}$ is the time-reversal operator $\hat{\Theta} = ie^{-i\pi\sigma_{y}}\hat{K}$, where $\hat{K}$ is the complex conjugation operator.
In the right part of Fig. (\ref{fig:fig2}) a non-zero $h_{z}\neq 0$ opens up a gap equal to $2\vert h_{z} \vert$ at the valence and conduction bands degeneracy points. Now the ferromagnetic order breaks the symmetry between the sublattices.
Spinors corresponding to these eigenvalues can be obtained analytically. We found it convenient to first change the basis to the one corresponding to $\left[ \sigma_{x}s_{y} - \sigma_{y}s_{x}\right]\phi_{\pm} = \pm \sqrt{s_{x}^2+s_{y}^2}\phi_{\pm}$ linear equation. For that we use unitary matrix which rotates to this basis,
\begin{align}
\hat{T}_{\bf k} = \frac{1}{\sqrt{2}}\left[\begin{array}{cc} e^{i\phi_{\bf k}} & e^{i\phi_{\bf k}} \\ 1 & -1 \end{array}\right], ~~
\hat{T}^{-1}_{\bf k} = \frac{1}{\sqrt{2}}\left[\begin{array}{cc} e^{-i\phi_{\bf k}} & 1 \\ e^{-i\phi_{\bf k}} & -1 \end{array}\right],
\end{align}
where $\phi_{\bf k} = \arctan\left( \frac{s_{x}}{s_{y}} \right)$ is the phase. Indeed, it can be verified that $\hat{T}^{-1}_{\bf k} \left[ \sigma_{x}s_{y} - \sigma_{y}s_{x}\right]\hat{T}_{\bf k} = \sqrt{s_{x}^2+s_{y}^2}\sigma_{z} \equiv s_{\bf k} \sigma_{z}$, as well as 
$
\hat{T}^{-1}_{\bf k} \sigma_{z} \hat{T}_{\bf k} = \sigma_{x}.
$
Then, the Hamiltonian of the system gets rotated as
\begin{align}
\hat{T}^{-1}_{\bf k} \hat{H} \hat{T}_{\bf k}
=
\left[\begin{array}{cc}  h_{z}\sigma_{x} -\mu & \xi_{\bf k} + i\gamma s_{\bf k} \sigma_{z} \\ 
 \xi_{\bf k} - i\gamma s_{\bf k} \sigma_{z} & h_{z}\sigma_{x} -\mu
\end{array} \right],
\end{align}
where $\hat{T}_{\bf k}$ comes with an identity matrix in the space of sublattices when acting on the full Hamiltonian.
Using symmetry properties of the rotated Hamiltonian, the eigenvectors of $\hat{T}^{-1}_{\bf k}\hat{H}\hat{T}_{\bf k} \psi_{\bf k} = \epsilon_{\bf k} \psi_{\bf k}$ equation are obtained in the following convenient form,
\begin{align}
\psi_{{\bf k};1;\pm} = \frac{1}{2}\left[ \begin{array}{c} \pm e^{i\frac{\chi_{{\bf k};1}}{2}} \\ e^{-i\frac{\chi_{{\bf k};1}}{2}} \\ e^{-i\frac{\chi_{{\bf k};1}}{2}} \\  \pm e^{i\frac{\chi_{{\bf k};1}}{2}}\end{array}\right],~
\psi_{{\bf k};2;\pm} = \frac{1}{2}\left[ \begin{array}{c} \pm e^{-i\frac{\chi_{{\bf k};2}}{2}} \\ e^{i\frac{\chi_{{\bf k};2}}{2}} \\ -e^{i\frac{\chi_{{\bf k};2}}{2}} \\  \mp e^{-i\frac{\chi_{{\bf k};2}}{2}}\end{array}\right],\label{beforeRotation}
\end{align}
where $\chi_{{\bf k};1/2} = \arctan\left( \frac{\gamma s_{\bf k}}{h_{z} \pm \xi_{\bf k}}  \right)$, and where $\psi_{{\bf k};1;\pm} $ spinor corresponds to $\epsilon_{{\bf k};1; \pm} $ dispersion, and $\psi_{{\bf k};2;\pm} $ to 
$\epsilon_{{\bf k};2; \pm}$. Original spinors are obtained from $\Psi_{\bf k} = \hat{T}_{\bf k}\psi_{\bf k}$ and read as
\begin{align}
\Psi_{{\bf k};1;+} = \frac{1}{\sqrt{2}}\left[ \begin{array}{c}  e^{i\phi_{\bf k}}c_{{\bf k};1} \\ i s_{{\bf k};1} \\ e^{i\phi_{\bf k}}c_{{\bf k};1}  \\  -i s_{{\bf k};1}\end{array}\right], 
\Psi_{{\bf k};1;-} = \frac{1}{\sqrt{2}}\left[ \begin{array}{c} -ie^{i\phi_{\bf k}}s_{{\bf k};1}\\ -c_{{\bf k};1}\\ -ie^{i\phi_{\bf k}}s_{{\bf k};1} \\   c_{{\bf k};1}\end{array}\right],\label{afterRotation1}
\end{align}
and
\begin{align}
\Psi_{{\bf k};2;+} = \frac{1}{\sqrt{2}}\left[ \begin{array}{c}  e^{i\phi_{\bf k}}c_{{\bf k};2}\\ - i s_{{\bf k};2}\\ - e^{i\phi_{\bf k}}c_{{\bf k};2} \\  - i s_{{\bf k};2}\end{array}\right], 
\Psi_{{\bf k};2;-} = \frac{1}{\sqrt{2}}\left[ \begin{array}{c} i e^{i\phi_{\bf k}}s_{{\bf k};2}\\ - c_{{\bf k};2} \\ -ie^{i\phi_{\bf k}}s_{{\bf k};2}  \\  - c_{{\bf k};2}\end{array}\right], \label{afterRotation2}
\end{align}
where we have defined $c_{{\bf k};1/2} = \cos\left( \frac{\chi_{{\bf k};1/2}}{2}\right)$ and $s_{{\bf k};1/2} = \sin\left( \frac{\chi_{{\bf k};1/2}}{2}\right)$ for brevity. It can be checked that these spinors form an othonormal set.

\begin{figure}[t] 
\begin{tabular}{cc}
\includegraphics[width=0.6 \columnwidth ]{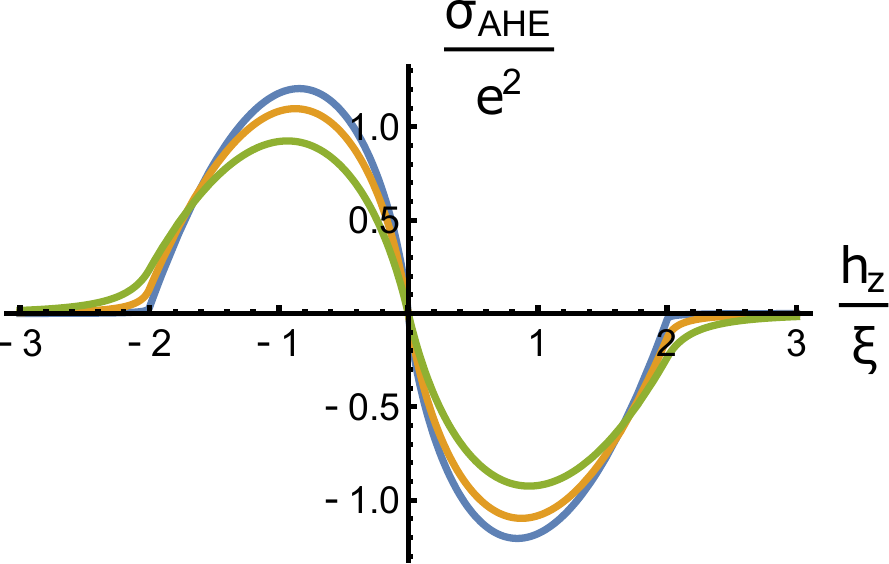} ~~~~ 
\end{tabular}

\protect\caption{Anomalous Hall effect conductivity $\sigma_{\mathrm{AHE}}$ plotted as a function of $h_{z}$ at $T=0$. Blue: $\gamma =0.1 \xi$, yellow $\gamma=0.3 \xi$, and green $\gamma=0.6 \xi$. It was assumed that $h=2\pi \hbar \equiv 1$. }
\label{fig:fig3}  
\end{figure}
We now wish to understand the structure of the anomalous Hall effect in the system.  
The anomalous Hall current ${\bf j}_{\mathrm{AHE}}$ in the applied electric field ${\bf E}$ is given by
\begin{align}\label{currentAHE}
{\bf j}_{\mathrm{AHE}} = \frac{e^2}{\hbar}
\left[ \int_{\mathrm{BZ}}\frac{d^2 k}{(2\pi)^2}\sum_{n = 1,2;m=\pm} {\bm \Omega}_{{\bf k};n;m} 
{\cal F}(\epsilon_{{\bf k};n;m} ) \right]\times {\bf E},
\end{align}
where ${\bm \Omega}_{{\bf k};n;m} $ is the Berry curvature, ${\cal F}(\epsilon)$ is the Fermi-Dirac distribution function, and the integration is over the Brillouin zone (BZ). We recall that $h=2\pi \hbar$ and set $h\equiv1$ in numerical calculations. In two dimensions the Berry curvature is along the normal to the plane of the system, i.e. ${\bm \Omega}_{{\bf k};n;m}  = \Omega_{{\bf k};n;m} {\bf e}_{z}$.
We are interested in the insulating case, hence everywhere below we set the Fermi energy to the gap, i.e. $\mu = 0$.
At zero temperature, $T=0$, it is enough to calculate the Berry curvature for valence bands only.
For $n=1,2$ we derive
\begin{align}
\Omega_{{\bf k}; n;-} &= \mathrm{Im}\left[  \partial_{x}\Psi^{\dag}_{{\bf k};n;-}  \partial_{y}\Psi_{{\bf k};n;-}  -   \partial_{y}\Psi^{\dag}_{{\bf k};n;-}  \partial_{x}\Psi_{{\bf k};n;-}  \right]
\nonumber
\\
&
=\frac{1}{2} \sin(\chi_{{\bf k};n})\left[ (\partial_{x}\phi_{\bf k})(\partial_{y}\chi_{{\bf k};n}) -  (\partial_{y}\phi_{\bf k})(\partial_{x}\chi_{{\bf k};n}) \right],
\end{align}
where $c_{x/y} = \cos(k_{x/y})$ is introduced for brevity, and $\partial_{x/y}$ is the partial derivative over $k_{x/y}$. Derivatives are $\partial_{x}\phi_{\bf k} = c_{x}s_{y}/s_{\bf k}^2$, $\partial_{y}\phi_{\bf k} = - c_{y}s_{x}/s_{\bf k}^2$, and 
\begin{align}
\partial_{x/y}\chi_{{\bf k};1/2} = \frac{\gamma s_{x/y} \left[ c_{x/y} (h_{z} \pm \xi_{\bf k})  + \xi s_{\bf k}^2 \right]}{s_{\bf k}\left[ (h_{z}\pm\xi_{\bf k})^2 +  \gamma^2 s_{\bf k}^2 \right]},
\end{align}
where note $\xi$ rather than $\xi_{\bf k}$ in the last term in the numerator.

\begin{figure}[h] 

\includegraphics[width=0.6 \columnwidth ]{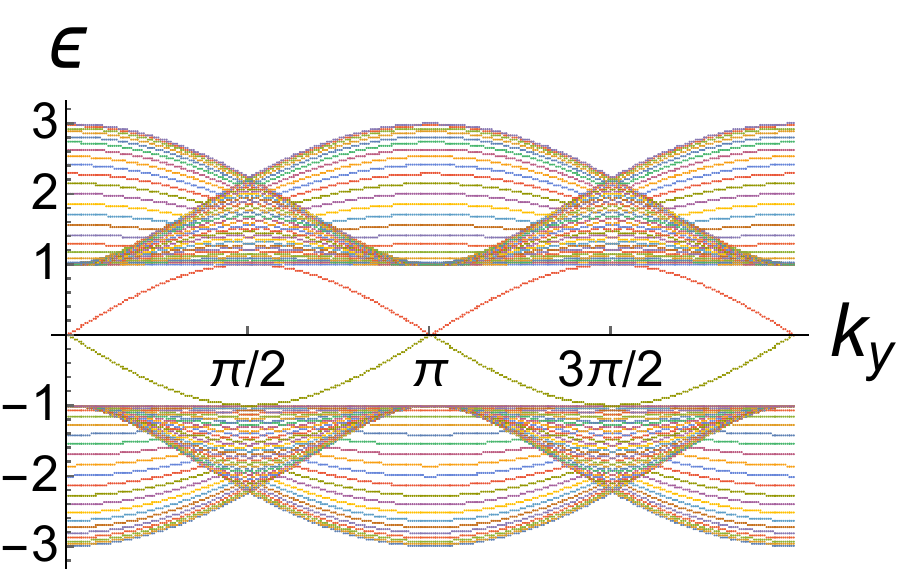} ~~~~

\protect\caption{
Spectrum of fermions numerically calculated in a strip geometry. Parameters are chosen to be $\xi = \gamma = h_{z} = 1$ for simplicty and in order to highlight the edge states in the gap. The edge states are there as long as $h_{z}\neq 0$ and $\vert h_{z}\vert<2\xi$.}
\label{fig:fig4}  
\end{figure}
\begin{figure}[t] 

\begin{tabular}{c}
\includegraphics[width=0.4 \columnwidth ]{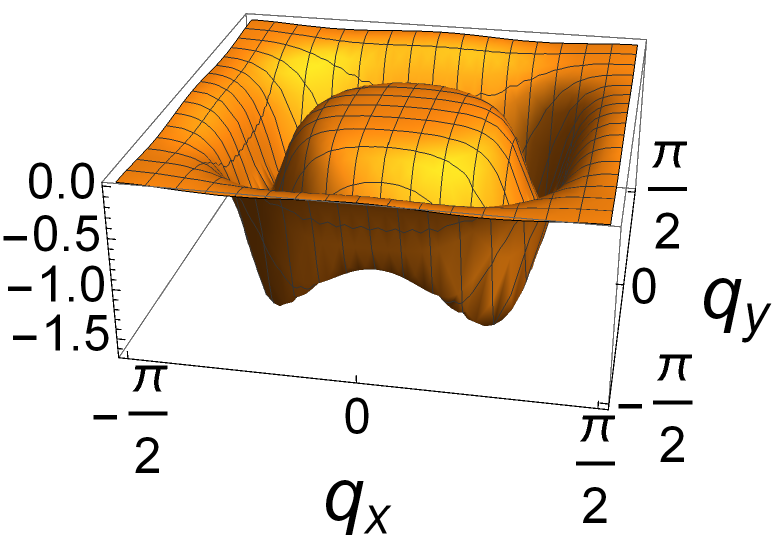} ~~
\includegraphics[width=0.4 \columnwidth ]{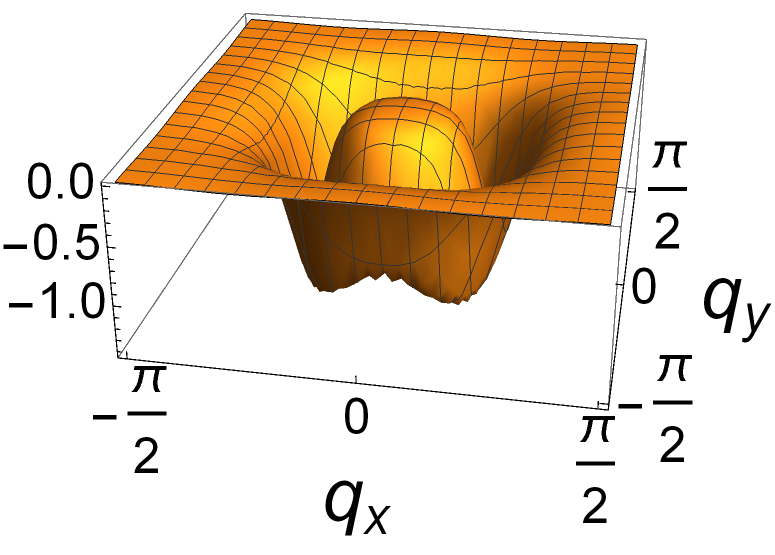} \\
\includegraphics[width=0.4 \columnwidth ]{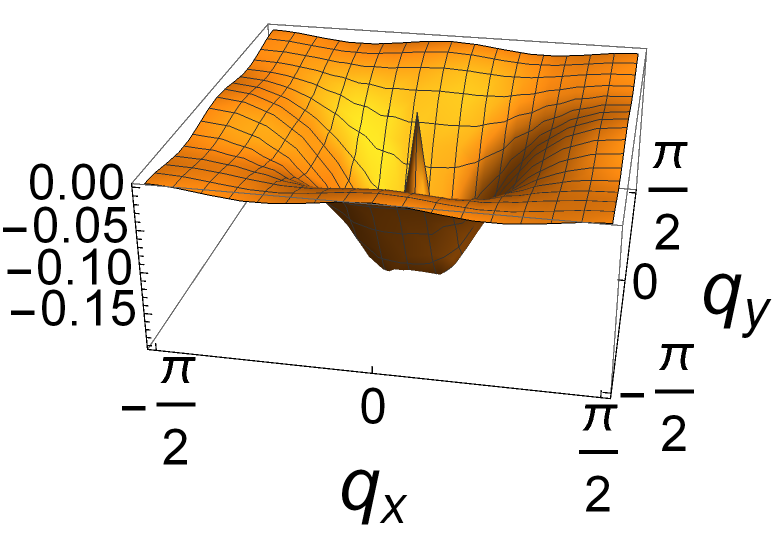} ~~
\includegraphics[width=0.4 \columnwidth ]{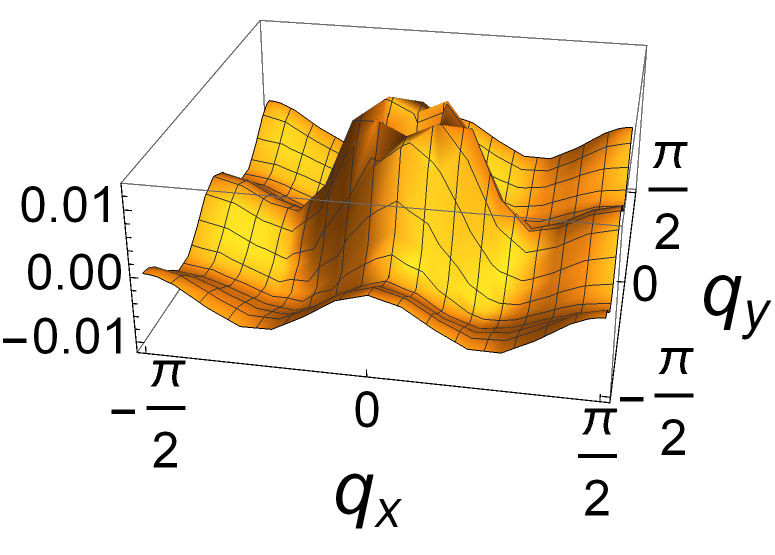} 
\end{tabular}

\protect\caption{
Evolution of the total Berry curvature of the valence band, $\Omega_{{\bf k}; 1;-} +\Omega_{{\bf k}; 2;-}$, with changing $h_{z}$ for fixed $\gamma=0.3 \xi$.
Top row: left for $h_{z}=\xi$ and right for $h_{z}=1.5\xi$. Bottom row: left for $h_{z}=2\xi$ and right for $h_{z}=2.5\xi$. Coordinates are defined as $q_{x/y}=\frac{1}{2}(k_{x}\pm k_{y})$ in terms of $k_{x/y}$ used throughout the text.}
\label{fig:fig5}  
\end{figure}

Thus, we finally obtain
\begin{align}
\Omega_{{\bf k}; 1/2;-}=\frac{\gamma^2 s_{\bf k}^2\left[ c_{x}c_{y}h_{z} \pm \xi_{\bf k} \right] }{2\left[ (h_{z} \pm \xi_{\bf k})^2 + \gamma^2 s_{\bf k}^2\right]^{\frac{3}{2}}},
\end{align}
where in deriving it we have used a definition of $\xi_{\bf k}$ and $c_{x/y}^2+s_{x/y}^2 = 1$ identity.
At $h_{z} = 0$ we have $\Omega_{{\bf k}; 1;-} +\Omega_{{\bf k}; 2;-} = 0$, and there is no anomalous Hall effect in the absense of the ferromagnetic order. We plot anomalous Hall conductivity $\sigma_{\mathrm{AHE}}/e^2$ in Fig. (\ref{fig:fig3}) as a function of $h_{z}$ for different values of $\gamma$. In addition, in Fig. (\ref{fig:fig4}) we plotted eigenvalues of fermions calculated in a strip geometry in $x$ coordinate with a hard-wall boundary condition. There are chiral edge states dispersing in the gap of the bulk spectrum. We checked that the edge states remain gapless in the parameter window of $\vert h_{z} \vert < 2\xi$. The edge states disappear as $h_{z}$ increases, namely, when it is $\vert h_{z} \vert \geq 2\xi$. Evolution of the total Berry curvature, $\Omega_{{\bf k}; 1;-} +\Omega_{{\bf k}; 2;-}$, of the valence band fermions for different values of $h_{z}$ is shown in Fig. (\ref{fig:fig5}). For example, at $h_{z}=2.5\xi$ the amount of positive and negative parts of the curvature becomes about equal, and hence the anomalous Hall conductivity indeed vanishes.
This is consistent with Fig. (\ref{fig:fig3}) and with the absence of the chiral edge states at $\vert h_{z}\vert > 2\xi$. We note that according to Fig. (\ref{fig:fig3}) the anomalous Hall conductivity isn't quantized in the region $\vert h_{z} \vert < 2\xi$. This is the region where there are chiral edge states, and the Berry curvature is mostly of the same sign for all $k_{x}$ and $k_{y}$. Absence of quantization is due to the parameter-dependent structure of the total Berry curvature for the valence band fermions shown in Fig. (\ref{fig:fig5}). 
Typically, quantization of Hall conductivity would mean that the integral of the Berry curvature over the BZ encircles half of the sphere on which the spin is defined. In that way the phase acquired by the spin is $\frac{\pi}{2}$. This is achieved by the spinor structure of the wave functions in which the angles $\phi_{\bf k}$ and $\chi_{\bf k}$ that parametrize the spin ( in our case they are defined in Eqs. (\ref{afterRotation1}) and (\ref{afterRotation2})), sweep corresponding surface on the unit sphere during the integration. This, for example, happens in the known theoretical model of a two-dimensional fermion gas with Rashba spin-orbit coupling and Zeeman magnetic field \cite{CulcerMacDonaldNiuPRB2003}. In our system, we also observe similar structure of the spinors given in Eqs. (\ref{afterRotation1}) and (\ref{afterRotation2}), with the only difference that the effective gap parameter that enters the angles $\chi_{{\bf k};12}$, i.e. $h_{z}\pm \xi_{\bf k}$ defined in the text after Eq. (\ref{beforeRotation}), is itself dependent on the momentum and even have nodes. The trajectories of the spin on the unit sphere in the course of the integraion no longer lead to the quantization.

\section{Conclusions}
To conclude, in this paper we have presented a theoretical model of ferromagnet coexisting with the noncollinear antiferromagnetic ordered. 
Interaction of conducting fermions with the noncollinear antiferromagnetic order is shown to result in the momentum-dependent exchange interaction of the former. The interaction is shown to be odd in momentum Eq. (\ref{antitoroidal}), and it does resemble standard Rashba spin-orbit coupling but with the difference that it breaks the time-reversal symmetry. We showed that the anomalous Hall effect is obatined in the system without any spin-orbit coupling. 
The anomalous Hall effect is shown to be consistent with the existence of chiral edge states. 
Proposed in this paper model allows to understand physics of anomalous Hall effect analytically \cite{VolkovPankratov1985,PankratovPakhomovVolkov1987,Haldane1988,CulcerMacDonaldNiuPRB2003}, and we hope it will receive further attention. 
Proposed in this paper mechanism of the anomalous Hall effect can be relevant to ferromagnets with a spontaneous magnetization coexisting with noncollinear antiferromagnetic order \cite{Turov1965,BorovikRomanov1988,Mineev1996,ChenNiuMacDonaldPRL2014,Liu2022,Fukami2025,Parkin2025}.

\section{Acknowledgements}
The authors thanks S.V. Aksenov, V.P. Mineev, M.S. Shustin, and P.O. Sukhachov for helpful discussions.
The author is grateful to Pirinem School of Theoretical Physics for hospitality during the Summer of 2025. 
This work is supported by FFWR-2024-0016.

\end{document}